\documentclass[twocolumn,trackchanges]{aastex7}

\newcommand{\siiv}{Si {\sc iv}}

\begin{document}

\title{Two-stage primary acceleration in filament initial eruption under a fan-spine magnetic configuration}

\author[0000-0001-6024-8399]{Haitang Li}
\affiliation{School of Physical Science and Technology, Southwest Jiaotong University, Chengdu 611756, People$'$s Republic of China}
\affiliation{Astrophysical Center, Southwest Jiaotong University, Chengdu 611756, People$'$s Republic of China}
\affiliation{Key Laboratory of Modern Astronomy and Astrophysics (Nanjing University), Ministry of Education, Nanjing 210093, People$'$s Republic of China}
\email[show]{lihaitang@swjtu.edu.cn; lyu@swjtu.edu.cn}

\author[0000-0002-1134-4023]{Ke Yu}
\affiliation{College of Physics and Electronic Engineering, Sichuan Normal University, Chengdu 610068, People$'$s Republic of China}
\email[]{}

\author[0009-0009-1911-399X]{Chang Zhou}
\affiliation{Key Laboratory of Modern Astronomy and Astrophysics (Nanjing University), Ministry of Education, Nanjing 210093, People$'$s Republic of China}
\affiliation{School of Astronomy and Space Science, Nanjing University, Nanjing 210093, People$'$s Republic of China}
\email[]{}

\author[0000-0002-8500-673X]{Qiang Liu}
\affiliation{School of Physical Science and Technology, Southwest Jiaotong University, Chengdu 611756, People$'$s Republic of China}
\affiliation{Astrophysical Center, Southwest Jiaotong University, Chengdu 611756, People$'$s Republic of China}
\email[]{}

\author[0000-0003-2837-7136]{Xin Cheng}
\affiliation{Key Laboratory of Modern Astronomy and Astrophysics (Nanjing University), Ministry of Education, Nanjing 210093, People$'$s Republic of China}
\affiliation{School of Astronomy and Space Science, Nanjing University, Nanjing 210093, People$'$s Republic of China}
\email[]{}

\author[0000-0002-4205-5566]{Jinhan Guo}
\affiliation{Key Laboratory of Modern Astronomy and Astrophysics (Nanjing University), Ministry of Education, Nanjing 210093, People$'$s Republic of China}
\affiliation{School of Astronomy and Space Science, Nanjing University, Nanjing 210093, People$'$s Republic of China}
\email[]{}

\author[0000-0002-2995-070X]{Feiyang Sha}
\affiliation{School of Physical Science and Technology, Southwest Jiaotong University, Chengdu 611756, People$'$s Republic of China}
\affiliation{Astrophysical Center, Southwest Jiaotong University, Chengdu 611756, People$'$s Republic of China}
\email[]{}

\author[0000-0002-1190-0173]{Ye Qiu}
\affiliation{Institute of Science and Technology for Deep Space Exploration, Suzhou Campus, Nanjing University, Suzhou 215163, China}
\email[]{}

\author[0000-0002-7694-2454]{Yu Liu}
\affiliation{School of Physical Science and Technology, Southwest Jiaotong University, Chengdu 611756, People$'$s Republic of China}
\affiliation{Astrophysical Center, Southwest Jiaotong University, Chengdu 611756, People$'$s Republic of China}
\email[]{}

\begin{abstract}
Understaning the filament rising process is crucial for unveiling the triggering mechanisms of the coronal mass ejections and forecasting the space weather. In this paper, we present a detailed study on the filament initial eruption under a fan-spine structure. It was found that the filament underwent two distinct acceleration stages corresponding to a calss M1.0 and M4.6 flare event, respectively. The first acceleration stage commenced with the filament splitting, after which the upper portion was subsequently heated being a hot channel and slow rose at an average speed of 22 km s$^{-1}$. A set of hot reverse C-shaped loops appeared repeatedly during the filament splitting and a hook structure was recognized at this phase, suggesting ongoing growth of the magnetic flux rope (MFR). When it reached a certain altitude, the hot channel appeared to get into a quasi-static phase with its upper edge seriously decelerated and lower edge expanding downward. Approximately 30 minutes later, as a distinct annular ribbon appeared outside the hook structure, the hot channel rose again at a velocity over 50 km s$^{-1}$ accompanied with rapid footpoints drifting, and experienced the second acceleration stage with its axial flux increased to $1.1 \times 10^{21}$ Mx. It is deduced that the filament initial eruption under a magnetic dome possess multi kinetic process. We suggest that the magnetic reconnection taken place within and beneath the filament continues to trigger the growth of pre-eruptive MFR and the first acceleration, when the magnetic reconnection above the filament plays a key role in the second acceleration. 
\end{abstract}

\keywords{\uat{Solar filaments}{1495};  \uat{Solar flares}{1496}; \uat{Solar magnetic fields}{1503}}

\section{Introduction} \label{sec1}
Coronal mass ejections (CMEs) are among the most violent eruptive phenomena in the solar atmosphere, capable of instantaneously releasing billions of tons of magnetized plasma into interplanetary space and potentially triggering significant space weather disturbances \citep{ChenP.F._2011_LRSP, Schmieder_2015_SoPh}. According to the statistical study by \cite{Gopalswamy_2003_ApJ}, over 70\% of prominence eruptions are associated with CMEs, indicating filaments or prominences are one of the primary candidates for the origins of CMEs \citep{Jingju_2004_ApJ, McCauley_2015_SoPh}. Therefore, unraveling the initiation processes of filament eruptions plays a crucial guiding role in effectively forecasting space weather.

Solar filaments are cold and dense plasma suspended in the solar corona \citep{Tandberg_1974_Book, Engvold_2015_ASSL}. They are believed to be trapped in magnetic dips above the polarity inversion lines (PILs) of the magnetic field at the photosphere \citep{Parenti_2014_LRSP}, where the magnetic field exists in the form of sheared arcades \citep{Kippenhahn_1957_ZA, Karpen_2001_ApJ}, or a magnetic flux rope (MFR) which is characterized as a bundle of twisted magnetic field lines wrapping around a central axis \citep{Kuperus_1974_A&A, Aulanier_1998_A&A_2, Chengxin_2014_ApJL, Lihaitang_2025_ApJL}. According to extensive observational studies and numerical simulations \citep{Green_2009_ApJ, Aulanier_2010_ApJ, Chengxin_2013_ApJL, Guoyang_2013_ApJ, Jiangchaowei_2018_ApJ, Jiangchaowei_2021_FrP}, MFRs have been regarded as an essential pre-eruptive structure for CMEs, which are usually formed before or during the solar eruption and manifest as sigmoids, hot channels and so on \citep{Hudson_1998_GeoRL, Green_2007_SoPh, Zhangjie_2012_NatCo, Chengxin_2013_ApJ}.

The essence of filament eruptions lies in the rapid release of non-potential magnetic energy accumulated in the corona and its conversion into plasma thermal and kinetic energy \citep{Moore_1980_Book, Masuda_1994_Natur}. Generally, the factors triggering filament eruption and facilitating energy release can be typically classified into magnetic reconnection and ideal magnetohydrodynamic (MHD) instabilities. The former mainly alters the topological connectivity of the magnetic field, thereby reducing the background constraints and causing the filament rise, e.g., tether-cutting reconnection, magnetic flux emergence and breakout model \citep{Chifor_2007_A&A, Moore_2001_ApJ, ChenP.F._2000_ApJ, Antiochos_1999_ApJ_1}. The mechanisms driving magnetic reconnection are often attributed to the photospheric shear and converging motions, which typically lead to magnetic flux cancellation on the photosphere \citep{VanBallegooijen_1989_ApJ, Yardley_2016_ApJ, Yardley_2018_ApJ, Priest_2018_ApJL}. However, the ideal MHD instabilities posit the existence of a critical threshold of the magnetic system$'$s intrinsic properties affecting the mechanical equilibrium of the entire MFR, primarily exemplified by the torus instability and kink instability \citep{ Torok_2004_A&A, Torok_2005_ApJ, Klim_2006_PhRvL, Aulanier_2010_ApJ}.

Although numerous triggering mechanisms have been proposed theoretically, it remains observationally challenging to unambiguously distinguish between them while the kinematic evolution of the MFR is highly synchronized with the soft X-ray flux variations during filament eruptions \citep{Zhangjie_2001_ApJ}. This temporal correspondence suggests that there are three distinct phases including a quasi-static phase, a slow rise phase and an impulsive erution phase in the early process of CMEs \citep{Zhangjie_2006_ApJ, Chengxin_2020_ApJ}. Notably, different physical mechanisms may dominate in each evolutionary stage \citep{Reeves_2015_ApJ, Joshi_2017_ApJ, Xingyaoyu_2024_MNRAS, Xuzhe_2024_MNRAS}. Through investigating the eruption of a hot channel, \cite{Chengxin_2023_ApJL} recently revealed that the slow rise of MFR began with the pre-flare magnetic reconnection which take place at a hyperbolic flux tube (HFT). Combining numerical simulations, \cite{Xingchen_2024_ApJ, Xingchen_2025_ApJ} further demonstrated that filament drainage also plays a significant role in triggering the slow rise of MFR, with its later stage being co-driven by the HFT reconnection and torus instability. While this model successfully shed light on the slow rise phase in simple dipole configurations, the question remains how the kinematic evolution evolves in complex multipolar fields and whether additional physical processes become operational when substantial changes occur in the coronal magnetic topology.

In this paper, we perform a detailed investigation of a filament eruption underlying a fan-spine structure to illustrate the kinematic evolution of the filament and the interactions with surrounding magnetic structures during the initial process. The instruments and data we used are introduced in Section 2, and the main results are shown in Section 3, which is followed by conclusions and discussions in Section 4.

\section{Instruments and data} \label{sec2}
The filament eruption was primarily observed by the Atmospheric Imaging Assembly (AIA; \citealp{Lemen_2012_SoPh}) on board the Solar Dynamics Observatory (SDO; \citealp{Pesnell_2012_SoPh}), which provides 7 extreme-ultraviolet (EUV) and 2 ultraviolet (UV) images of the full solar disk with a pixel size of $\sim$ 0.6$^{''}$ and temporal cadence of $\sim$ 12 s and 24 s, respectively. Here, the images centered at 131 {\AA} ($\sim$ 10 MK) and 171 {\AA} ($\sim$ 0.6 MK) are used for detecting the pre-eruptive MFR (hot channel) and background coronal loops, the 211 {\AA} ($\sim$ 2 MK) images are used for detecting the associated eruption activities (e.g., filament, coronal dimming and post flare loop). The data from H$\alpha$ Imaging Spectrograph \citep{Liuqiang_2022_SCPMA} on board the Chinese H$\alpha$ Solar Explorer (CHASE; \citealp{Lichuan_2019_RAA, Lichuan_2022_SCPMA}) are also used to detect the filament, which provides the reconstructed full solar disk images with a spatial resolution of $\sim$ 0.52 per pixel and time cadence of $\sim$ 60 s at H$\alpha$ band (6562.8 $\pm$ 3.1 {\AA}).

Simultaneously, the Helioseismic and Magnetic Imager (HMI; \citealp{Scherrer_2012_SoPh, Schou_2012_SoPh}) on board SDO provides the line-of-sight magnetograms with a pixel size of $\sim$ 0.6$^{''}$ and time cadence of $\sim$ 45 s, and the vector magnetograms with a temporal resolution of $\sim$ 12 minutes. In this paper, the vector magnetograms from the Spaceweather HMI Active Region Patch \citep{bobra_2014} are used for reconstructing the three-dimensional coronal magnetic field.

The Interface Region Imaging Spectrograph (IRIS; \citealp{Pontieu_2014_Solphy}) provides high-resolution spectra and slit-jaw images (SJIs) via a slit to diagnose the eruption in lower atmosphere. The spectral resolutions are 0.025 {\AA} and 0.051 {\AA} in the far-ultraviolet (FUV; 1332--1358 {\AA}) and near-ultraviolet (NUV; 2783--2834 {\AA}, 1389--1407 {\AA}) wavelengths. The IRIS slit can be used in a sit-and-stare or raster scan mode with a width of 0.33$^{''}$. The SJIs have a field of view of $120^{''} \times 119^{''}$ at 4 passbands (1400, 1330, 2796 and 2832 {\AA}) with a time cadence of 19 s and a pixel scale of $0.166^{''}$. On 09-Aug-2024, IRIS observed the AR 13779 in the sit-and-stare mode with a time cadence of 9.4 s, which well covered the filament initial eruption. Here, we mainly focus on the \siiv~line at 1402.77 {\AA}, which forms in the transition region with a formation temperature of $\sim$ 0.1 MK and usually can be  treated as optically thin.  And the IRIS/SJI 1400 {\AA} images are used to analyse the flare ribbon evolution during the initial eruption. Generally, the IRIS data have been calibrated through dark current subtraction and wavelength, geometry and flat field correcttion. Since we are primarily concerned with the intensity and Doppler velocity of the \siiv~line at 1402.77 {\AA} profile, here, we just report the position of the line peak and utilize it to calculate the Doppler velocity. All images have been reprojected to the same reference time (09-Aug-2024 19:00:00 UT) for better comparison.
 
\section{Results} \label{sec3}
\subsection{Overview of the filament eruption}
\begin{figure*}[ht!]
	\centering 
	\includegraphics[width=0.9\textwidth,height=0.675\textwidth]{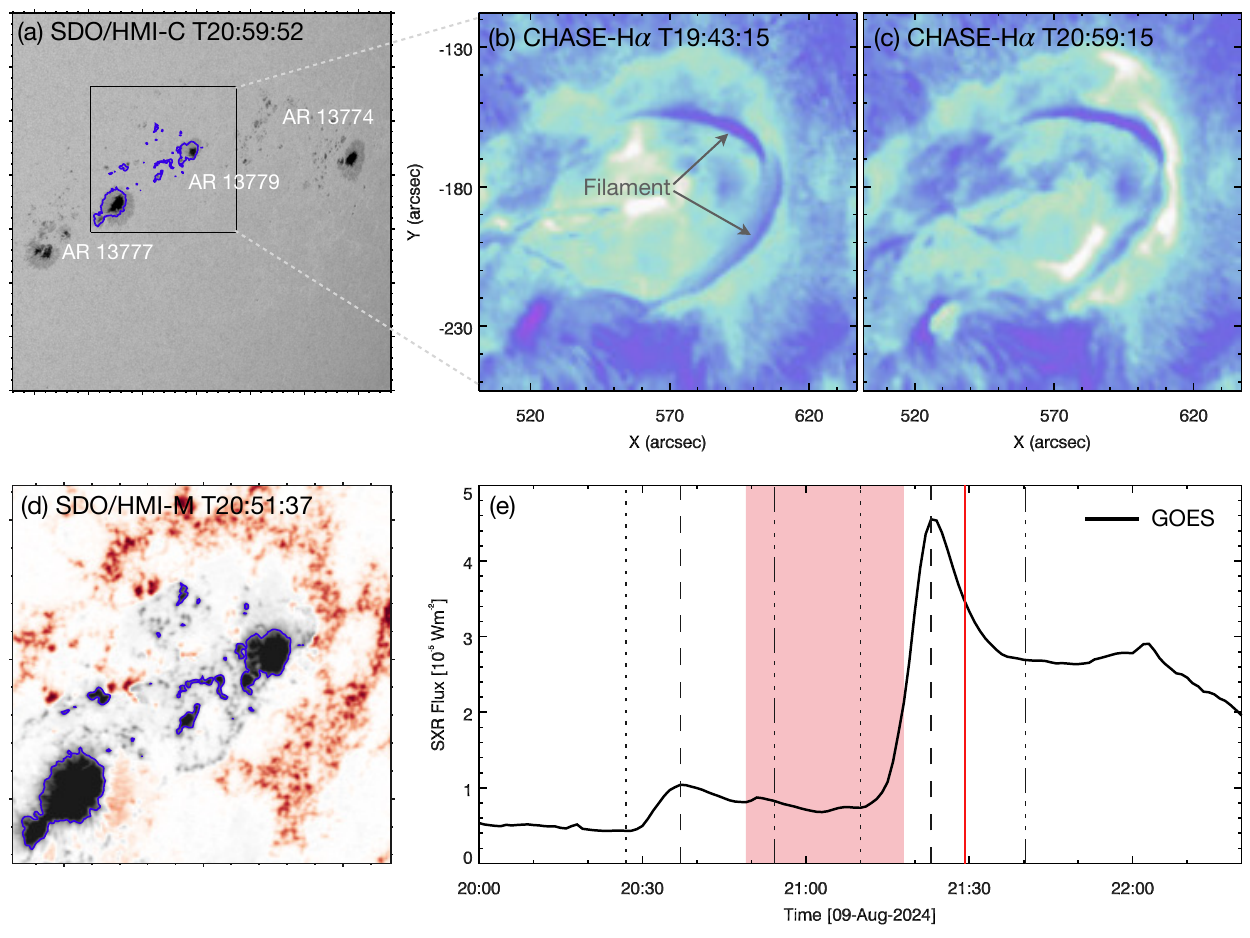}
	\caption{An overview of the filament eruption. (a) Photospheric continuum image of ARs observed by SDO/HMI. The black box outlines the field of view (FOV) for panels (b)-(d). (b)-(c) H$\alpha$ images showing the filament before (b) and during (c) the eruption observed by CHASE. (d) Line-of-sight magnetogram with the blue contours at B = - 500 G. The red and black regions represent positive and negative magnetic field, respectively. (e) Temporal evolution of GOES 1-8 {\AA} soft X-ray flux. The vertical dotted, dashed and dot-dashed lines indicate the start, peak and end time for each flare. The shadow region shows the operation period of CHASE and the solid red line indicates the end time of the IRIS observation.   \label{fig1}}
\end{figure*}

On 2024 Augest 09, a filament eruption was observed in NOAA active region (AR) 13779 accompanied by intense energy release, and produced a halo CME with its projected linear speed over 600 km s$^{-1}$. As shown in Figure \ref{fig1}(a) and \ref{fig1}(d), the AR was connected with another two NOAA AR 13777 and 13774 on the photosphere, where the sunsupt was domainated by negative magnetic filed and surrounded by dispersed positive poles, presenting a classical magnetic field with coronal fan-spine structure. From CHASE H$\alpha$ images, two filament segements were observed to be loacated over the mainly elliptic polarity inversion lines (PILs), mainifested as a reversed C-shaped filament at 19:43 UT (see Figure \ref{fig1}(b)). After 21:00 UT, the south filament went through a successful eruption with partial cool materials and north filament segement remained (see Figure \ref{fig1}(c)). On the other hand, the GOES 1-8 {\AA} soft X-ray (SXR) flux presented two apparent increase, as shown in Figure \ref{fig1}(e), the first of which started at $\sim$ 20:27 UT, peaked at $\sim$ 20:37 UT, ended at $\sim$ 20:54 UT, and the second started at $\sim$ 21:10 UT, peaked at $\sim$ 21:23 UT, ended at $\sim$ 21:40 UT, indicating two flare events took place during the filament eruption that reached M-class 1.0 and 4.6, respectively. Fourtunally, the initial process of the filament eruption was concurrently captured by CHASE and IRIS, providing a rare opportunity to study the early evolution of the filament eruption.

\subsection{Initial eruption of the filament}
\begin{figure*}[ht!]
	\centering 
	\includegraphics[width=0.9\textwidth,height=0.9\textwidth]{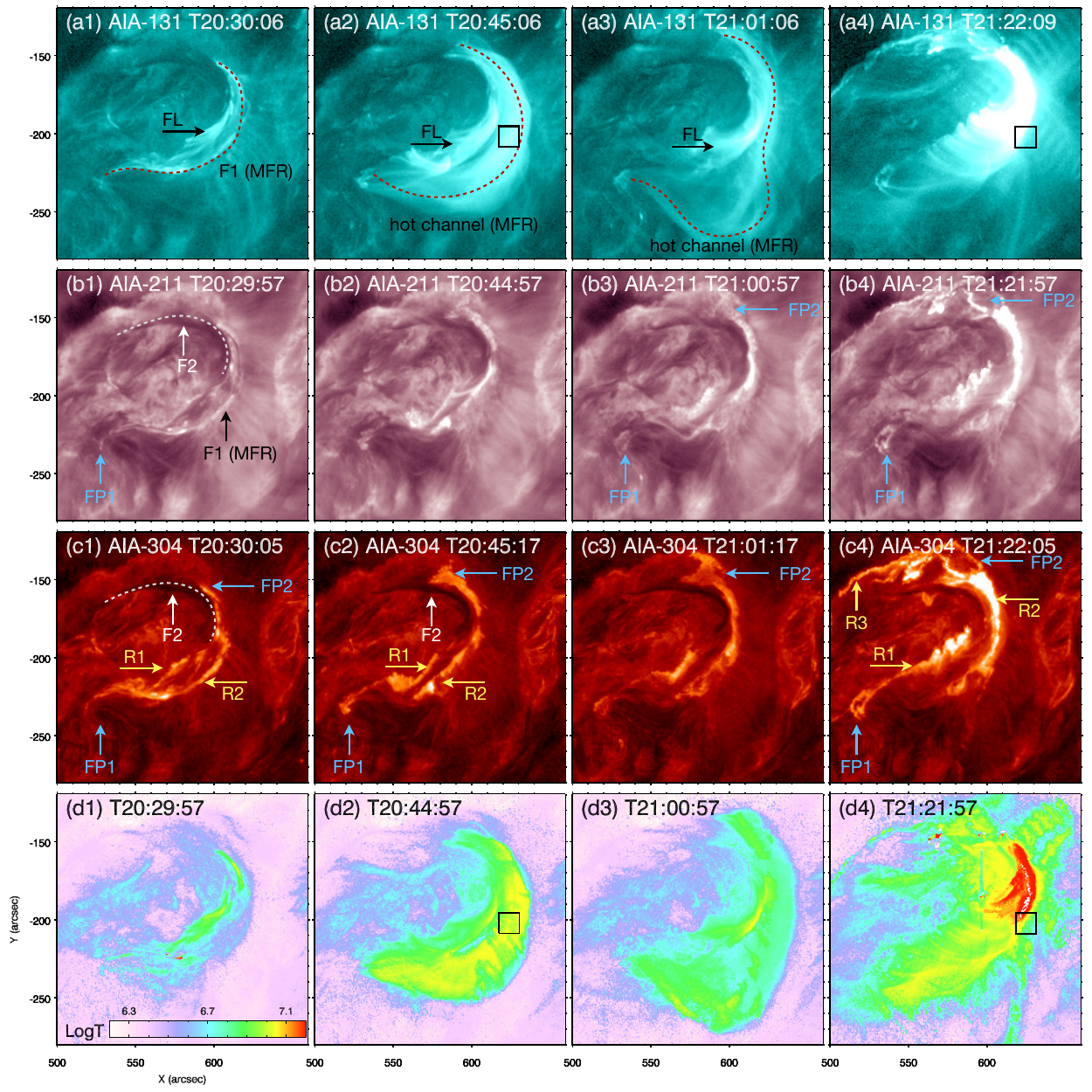}
	\caption{Images of the filament eruption. (a1)-(a4), (b1)-(b4) and (c1)-(c4) show the  filament eruption in AIA 131 {\AA}, 211 {\AA} and 304 {\AA} passband, respectively. The red dashed lines in panels (a1)-(a3) denote the erupted filament F1 and hot channel. The black arrows in panels (a1)-(a3) indicate the occurred flare loop (FL). The white dashed lines in panel (b1) and (c1) denote the remained filament F2.  The blue arrows in panels (b1)-(c4) indicate the foot points (FP1 and FP2) of the hot channel. The yellow arrows in panels (c1)-(c4) indicate the appeared flare ribbons. (d1)-(d4) show the average temperature derived by DEM method. The black box indicates the region to get the mean temperature of the hot channel in Figure \ref{fig4}. A 17 s animation including AIA 131 {\AA}, 211 {\AA} and 304 {\AA} is available online to present the filament splitting and eruption from 19:00 UT to 22:30 UT.} \label{fig2}
\end{figure*}

Over the several hours preceding the filament eruption, multiple instances of brightening were observed along the filament spine, as shown in Figure \ref{afig1} (also see the online movie in 171 {\AA} attached to Figure \ref{afig1}). The south filament gradually split into upper (here after called filament F1) and lower sections (filament F2), as indicated by the black and white arrows in Figure \ref{afig1}, implying the filament splitting was caused by the intermittent magnetic reconnection taken place in the filament. Coinciding with these brightenings, a set of inverse C-shaped loops underwent repeated heating in 94 {\AA} (see Figure \ref{afig2} and the online movie in 94 {\AA} attached to Figure \ref{afig1}), indicating the ongoing formation of a hot channel. At about 20:30 UT, the upper partial F1 was rapidly lifted and heated, forming a more coherent hot channel observed in AIA high-temperature passbands (see Figures \ref{fig2}(a2) and \ref{afig2}(f)), which is a typical indicator for MFR \citep{Zhangjie_2012_NatCo}. Simultaneously, a distinct set of flare loops (FL in Figures \ref{fig2}(a1)-\ref{fig2}(a3)) and flare ribbons (R1 and R2 in Figures \ref{fig2}(c1)-\ref{fig2}(c4)) were observed under the hot channel.

\begin{figure*}[ht!]
	\centering 
	\includegraphics[width=0.9\textwidth,height=0.9855\textwidth]{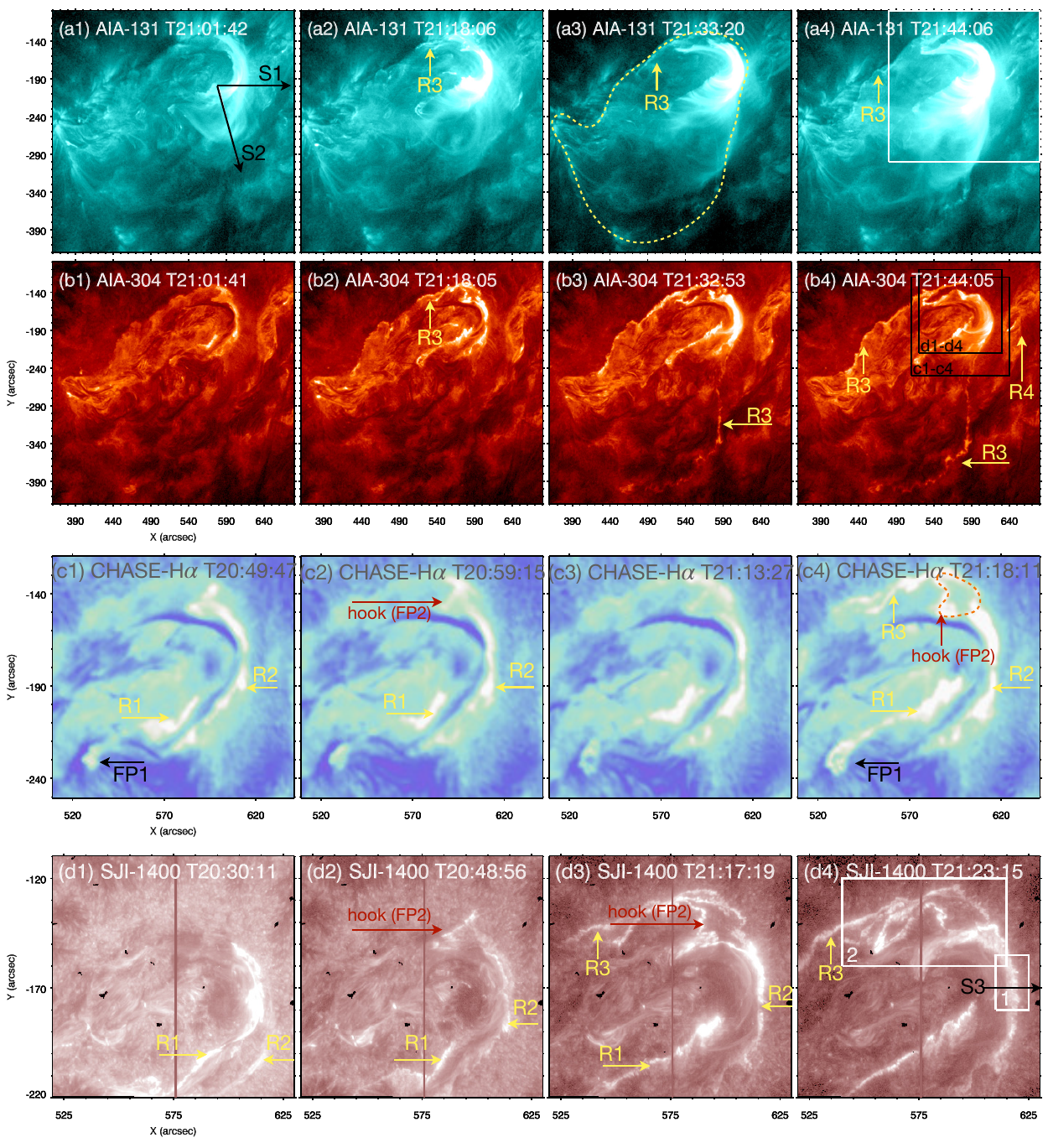}
	\caption{Evolution of the flare ribbons. (a1)-(a4) and (b1)-(b4) show the evolution of hot channel and flare ribbons in a larger filed of view (FOV) in AIA 131 {\AA} and 304 {\AA}. (c1)-(c4) and (d1)-(d4) show the evolution of flare ribbons observed by CHASE and IRIS. The yellow arrows denote the flare ribbon R1-R4 and the yellow dashed line contours the large circle ribbon R3. The black arrows in panels (a1) and (d4) indicate the slices S1-S3 to obtain the time-distance diagrams in Figure \ref{fig4}. The white boxes denote the integral region to get the light curves in Figure \ref{fig4}. The black boxes outline the FOV of panels (c1)-(c4) and (d1)-(d4), respectively. The black arrows in panels (c1) and (c4) indicate the south footpoint (FP1) of the filament F1. The red arrows indicate the identified hook structures. The orange dashed line contours the recongrized footpoint to integrate the axial flux of the MFR. A 10 s animation including AIA 131 {\AA}, 304 {\AA} from 20:26 UT to 22:30 UT and IRIS 1400 {\AA} from 19:52 UT to 21:28 UT is available online to present the evolution of the flare ribbon. \label{fig3}}
\end{figure*}

After the filament splitting, the upper portion F1 rose slowly with its north footpoint (FP2) drifted further northward, as shown in Figure \ref{fig2}(c1)-\ref{fig2}(c2) (also see the online movie attached to Figure \ref{fig2}), indicating continued growth of the pre-eruptive MFR. Subsequently, the upper edge of the hot channel reached to a nearly stationary platform showing a quasi-static phase, but its south leg kept expanding laterally as shown in Figures \ref{fig2}(a2)-\ref{fig2}(a3) and \ref{fig3}(a1)-\ref{fig3}(a2). During this phase, a clearly hook structure appeared at the north end of flare ribbon R2 at about 20:48 UT (see Figures \ref{fig2}(c2) and \ref{fig3}(d2)), representing the envelopment of MFR  footpoint FP2. Furthermore, the north footpoint (FP2) of the hot channel gradually swelled and drifted toward to northeast while the south footpoint (FP1) stayed almost fixed, as shown in Figures \ref{fig2}(c2)-\ref{fig2}(c4), suggesting the MFR was further growing up which can also be denoted by the enlarged hook structure in the end of flare ribbon R2 (see Figures \ref{fig3}(c1)-\ref{fig3}(c4) and \ref{fig3}(d1)-\ref{fig3}(d4)). As another distinct flare ribbon R3 appeared north of the hook at around 21:13 UT (see Figure \ref{fig3}(d3) and the online movie in 131, 304 {\AA} attached to Figure \ref{fig2}), the hot channel rapidly ascended and got into an impulsive rising phase, experiencing a successful ejection at approximately 21:20 UT companied with an M4.6 flare event (see Figure \ref{fig2}(a4)). And the north footpoint (FP2) of the hot channel rapidly drifted along the inner edge of the flare ribbon R3, forming a southward-concaved hook structure (see Figures \ref{fig2}(c4) and \ref{fig3}(c4)) in the impulsive phase.

Moreover, according to the differential emission measure (DEM) analysis \citep{Cheung_2015_ApJ}, we reconstructed the DEM with six coaligned AIA EUV images and then calculated the average temperature of the hot channel, which was derived to be approximately 8$\sim$15 MK during these two flare events (see Figure \ref{fig2}(d1)-\ref{fig2}(d4)). In addition, the temporal evolution of the mean temperature of the hot channel was presented in Figure \ref{fig4}(g).

\subsection{Temporal evolutions during the initial process}
\begin{figure*}[ht!]
	\centering 
	\includegraphics[width=0.9\textwidth,height=0.9\textwidth]{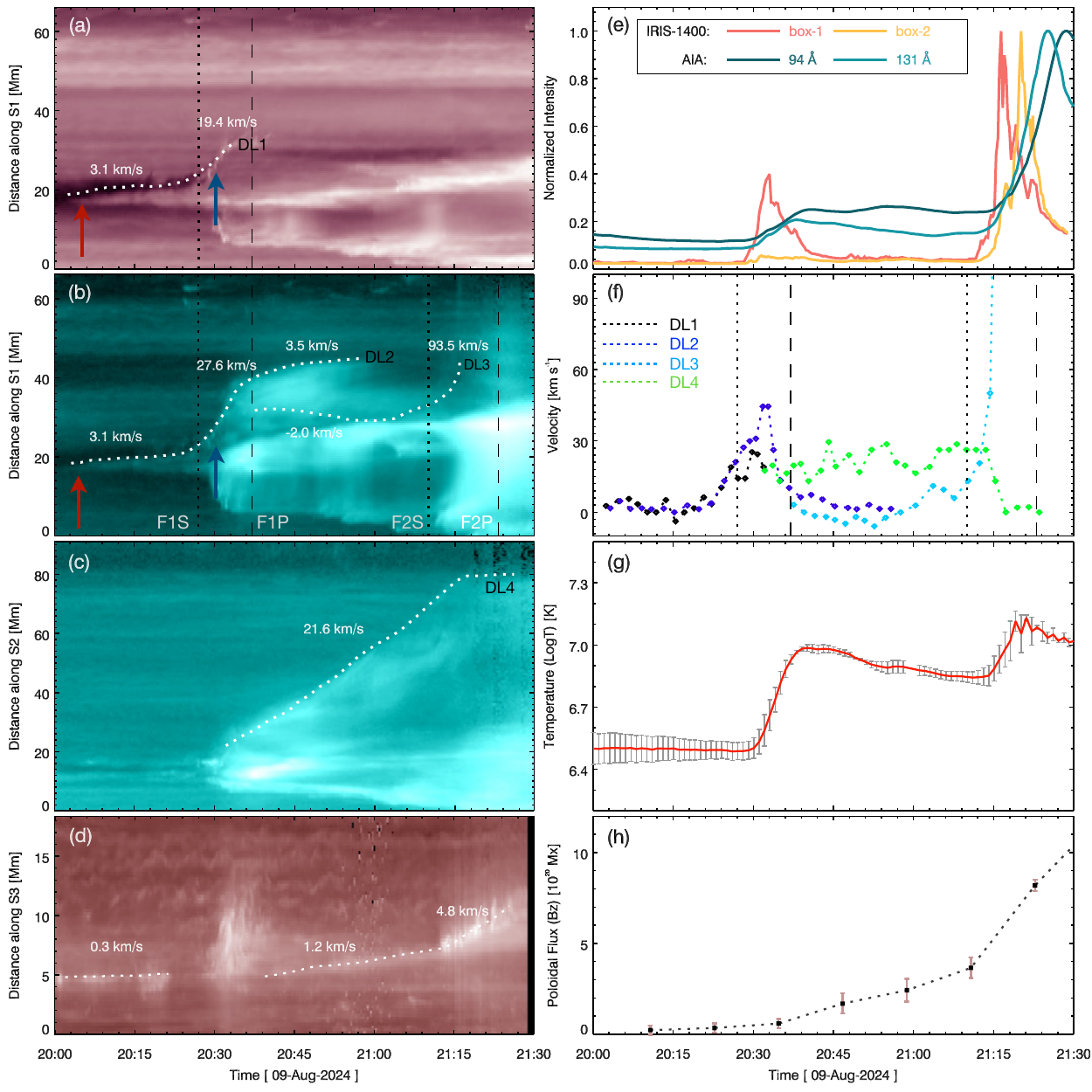}
	\caption{Dynamical evolution during the filament eruption. (a)-(d) Time-distance diagrams along the slice S1, S2 and S3 in Figure \ref{fig3}, respectively. (e) Normalized light curves at AIA 94, 131 {\AA} in the white box in Figure \ref{fig3}(a3), and IRIS 1400 {\AA} in box 1 and 2 in Figure \ref{fig3}(c4). (f) Velocity derived from the white dashed lines in panels (a)-(c). (g) Temporal evolution of the mean temperature of the hot channel as indicated by the black square in Figure \ref{fig2}. (h) Temporal evolution of the magnetic flux of the MFR. The vertical dotted and dashed lines denote the start and peak time of two flares, respectively. \label{fig4}}
\end{figure*}

To better illustrate the temporal evolution of the filament eruption, we constructed three slices (S1, S2, and S3) along the radial and lateral directions of the eruption and perpendicular to the flare ribbon R2, respectively, to generate time-distance diagrams depicting the hot channel rise process and flare ribbon separation, as shown in Figures \ref{fig4}(a)-(d). It was found that the splitting motion within the filament began at approximately 20:06 UT, as indicated by the red arrows in Figures \ref{fig4}(a) and \ref{fig4}(b). The upper portion (F1) started ascending at an average velocity in 3.1$\sim$27.6 km s$^{-1}$, forming a hot channel at around 20:30 UT (see the bule arrows in Figures \ref{fig4}(a) and \ref{fig4}(b)) and resulting in the first M1.0 class flare event. Concurrently, a pronounced intensity enhancement appeared beneath the filament at 1400 {\AA} (box-1 in Figure \ref{fig3}(c4)), with the light curves at 131 and 94 {\AA} also showed an increase (see Figure \ref{fig4}(e)). After the flare peaked, the lower edge of the hot channel continued descending at a velocity of 2 km s$^{-1}$, while the rising speed of the upper edge decelerated to 3.5 km s$^{-1}$ in a quasi state, suggesting the MFR was possibly constrained by the overlying background field. During this quasi-static phase, the south leg of the hot channel expanded laterally at a velocity of 21.6 km s$^{-1}$ (see Figures \ref{fig4}(b) and \ref{fig4}(c)). The light curves at 131 and 94 {\AA} entered a relatively stable plateau, with intensities increasing by approximately 20\%, while the 1400 {\AA} emition under the hot channel diminished to almost negligible levels. Meanwhile, the average temperature of the hot channel decreased slightly but stabilised at around 8 MK (see Figure \ref{fig4}(g)). This indicates that magnetic reconnection continued in the corona, maintaining MFR growth and providing gentle heating.

Within approximately half an hour, the hot channel abruptly accelerated upward at speeds ranging from 48.1 to 100 km s$^{-1}$ entering the impulsive rising phase. Eventually, the hot channel was successfully ejected at around 21:20 UT, producing the second M4.6 class flare accompanied by significant intensity enhancements across all EUV/UV passbands (see Figures \ref{fig4}(b) and \ref{fig4}(e)). The flare ribbon separated at a velocity of 4.8 km s$^{-1}$ during the impulsive rising phase, significantly faster than the separation speed of 0.3$\sim$1.2 km s$^{-1}$ in the quasi-static phase and the first slow rising phase. The average temperature of the hot channel which was heated to be around 10 and 15 MK in two flare events varied in synchronization with the SXR flux and AIA 131 {\AA} emission (see Figure \ref{fig4}(g)). Notably, the average temperature was derived from the region within the black box in Figure \ref{fig2}. Before approximately 21:13 UT, the hot channel was clearly separated from the underlying flare loops, meaning the temperature primarily represents that of the hot channel. However, after 21:13 UT, the hot channel was rapidly lifted, triggering the second flare. At this stage, the projection of the hot channel overlapped almost completely with the flare loops, and the intensity of the latter was significantly stronger. Consequently, the derived temperature in the black box now predominantly reflects the temperature of the flare loops, possibly exceeding 15 MK. Moreover, the time-distance diagram shows the south leg exhibited a stagnation after the hot channel erupted, suggesting the presence of a quasi-separatrix layer (QSL) in the magnetic field at that site.

Furthermore, the hook corresponding to the FP2 appears as a closed region in the H$\alpha$ band during the filament eruption, as shown in Figures \ref{fig3}(c2)-\ref{fig3}(c4). Under the higher resolution of IRIS observations, this hook also presents as a closed structure at sometimes (see Figures \ref{fig3}(d3)-\ref{fig3}(d4)). However, this hook is merged with the outer ribbon R3, making it difficult to distinguish between them. Nevertheless, combining the footpoints identified in AIA 211 and 131 {\AA} images, the dimming regions and the outer ribbon R3, we traced the hook structure at the terminus of flare ribbon R2 in IRIS 1400 {\AA} images to determine the north footpoint of the MFR, as marked by the orange dashed line in Figure \ref{fig3}(c4). Using HMI vector magnetograms, we then caculated the poloidal flux of the MFR (see Figure \ref{fig4}(h)), and find that the axial flux increased by $0.6 \times 10^{20}$ Mx during the first slow rising phase (from the filament initial splitting to the first flare peak). Subsequently, the poloidal flux further grew to $3.7 \times 10^{20}$ Mx at an average increasing velocity of $1.4 \times 10^{17}$ Mx s$^{-1}$ in the quasi-static plateau (from the first flare peak to the statr of second flare), then rapidly intensified to over $1.1 \times 10^{21}$ Mx ($5.7 \times 10^{17}$ Mx s$^{-1}$) in the impulsive rising phase (from start to peak of the second flare). These observations suggest that while the magnetic reconnection within the filament primarily drove the filament's ascent and formed the pre-eruptive MFR's core, the persistent moderate reconnection during the quasi-static phase was crucial for its subsequent development.  

\subsection{Magnetic reconnections above the filament}
\begin{figure*}[ht!]
	\centering 
	\includegraphics[width=0.9\textwidth,height=0.9\textwidth]{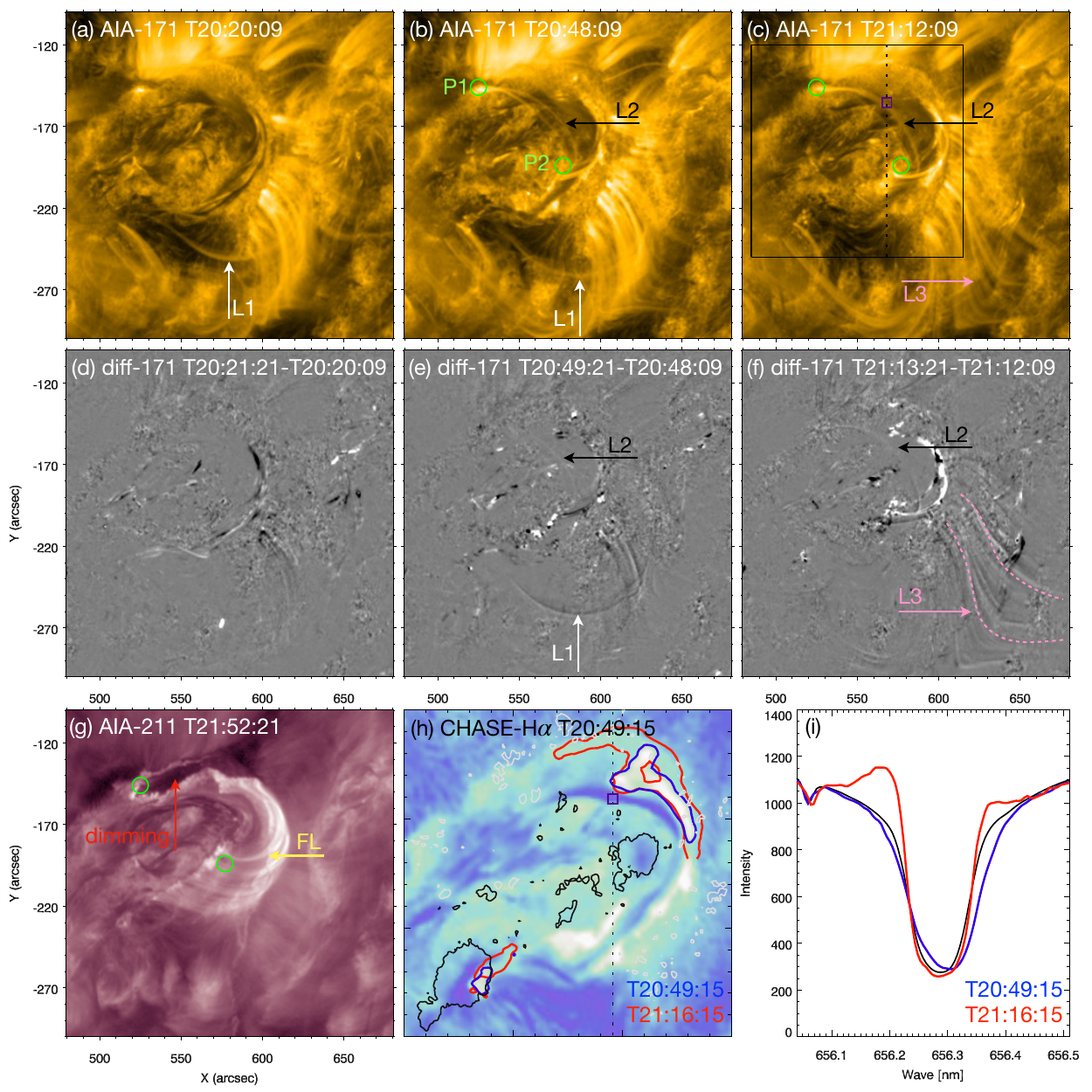}
	\caption{Configuration of coronal loops. (a)-(c) AIA 171 {\AA} images showing the changes of coronal loops. The white, black and pink arrows indicate different coronal loops L1, L2 and L3, respectively. The green circles indicates the footpoints of sheared arcade L2. The black box shows the FOV in panel (h). (d)-(f) Running difference images at AIA 171 {\AA}. The pink dashed lines show the opened coronal loop L3. (g) Footpoint drifting and coronal dimming at 211 {\AA}. The yellow arrow indicates the post flare loops. The red arrow indicates the dimming region. (h) The remained H$\alpha$ filament overlaid with magentic field.  The white and black contours represent the positive and negative magnetic polarities of $\pm$ 500 G. The blue and red contours represent the brightening edges at the footpoint. (i) H$\alpha$ profiles at the filament as pointed by the purple square in panel (c) and (h). \label{fig5}}
\end{figure*}

During the initial process, two types magnetic reconnections above the filament were identified as essential for the successful eruption and footpoint drifts. As shown in Figure \ref{fig5}(a), an arched loop (L1) and nearby curved coronal loops (L3) appeared above the south filament before the eruption. As the filament F1 rising up, the arched loop L1 expanded gradually upward (also see the online movie in 171 {\AA} attached to Figure \ref{afig1}). Moreover, a sheared loop (L2) appeared in the northwestern of the filament at approximately 20:48 UT, while the curved coronal loops L3 started moving progressively. During the quasi-static phase, the arched loop L1 further expanded and disappeared at around 21:13 UT. Meanwhile, the population of these curved loops L3 increased noticeably at around 21:13 UT, as presented in Figure \ref{fig5}(f), showing an observational trend of northward expansion. Simultaneously, the annular ribbon R3 appeared outside the north footpoint of the MFR (see Figure \ref{fig3}(d3)), which evolved into a larger circular ribbon R3 during the impulsive phase (see Figures \ref{fig3}(b3)-\ref{fig3}(b4) and the attached moive). Although the southeastern part of this circular ribbon R3 appears somewhat faint, the extrapolation of magnetic field reveals the presence of a null-point above the filament, yielding a clear fan-spine topology (see Figure \ref{fig7}(g)). Moreover, the QSL of fan footprints aligns well with the circular ribbon R3, as shown in Figure \ref{fig7}(h).  These collective observations suggest a scenario that the overlying loops L1 in Figure \ref{fig4}(a)-\ref{fig4}(b) above the south filament were highly forced by the continued growth of the MFR during the quasi-static phase. These arched loops then underwent a reconnection with the overlying background field in the QSL that associated with the null-point. This reconnection is similar to the breakout reconenction which sequentially reduces the downward magnetic pressure and allows more magnetic field lines to participate in the reconnection process. As a result, it would facilitate the subsequent rapid ascent of the hot channel, thereby generating the circular flare ribbon R3 and the nearby riboon R4 (see Figure \ref{fig3}(b4)).

\begin{figure*}[ht!]
	\centering 
	\includegraphics[width=0.9\textwidth,height=0.9\textwidth]{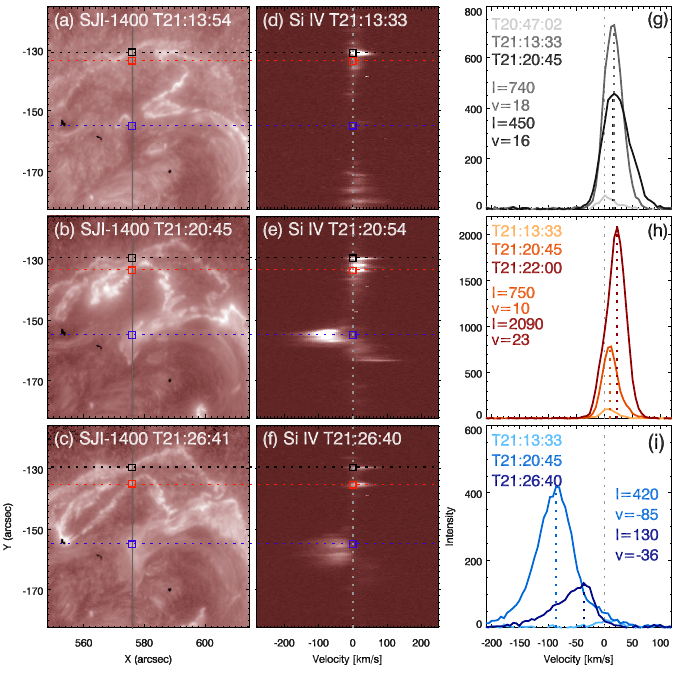}
	\caption{Spectral characteristics during the filament eruption. (a)-(c) IRIS 1400 {\AA} SJI images displaying the flare ribbons in the transition region. The black, red and blue dashed line denotes the location of flare ribbon R3, inner edge of the hook and remained filament, respectively. (d)-(f) Spectra of the \siiv 1402.77 {\AA} line at the slit as shown in panels (a)-(c). The vertical gray lines denote the slit of IRIS. The black, red and blue square indicates one pixel at the slit. (g)-(i) Profiles of the \siiv 1402.77 {\AA} line at R3, hook and filament. Quantities I and V correspond to the peak intensity and Doppler velocity.  \label{fig6}}
\end{figure*}

On the other hand, the north footpoint (FP2) of the MFR was observed to undergo rapid drifting, which was potentially attributed to the magnetic reconnection between the MFR and nearby arcades. During the quasi-static phase, the sheared loop L2 appeared east of the MFR footpoint (see Figures \ref{fig5}(b) and \ref{fig5}(c)). As the MFR entered the eruptive phase, its footpoint P1 gradually became obscured by dimming regions that associated with the MFR (see Figures \ref{fig5}(c) and \ref{fig5}(g)), indicating the occurrence of "ar-rf" reconnection \cite{Aulanier_2019_A&A}. Notably, while the north footpoint (FP2) of the MFR displayed rapid displacement with significant brightening, the south footpoint (FP1) that rooted in a sunspot remained stationary, as shown in Figures \ref{fig3}(c1)-\ref{fig3}(c4) and \ref{fig5}(h). This differential behavior demonstrates that footpoint drift depends on local magnetic field strength, occurring more readily in weak-field regions. Additionally, the underlying filament showed strong disturbances during the eruption, exhibiting prominent enhancement in H$\alpha$ blue-wing which was also detected by IRIS (see Figures \ref{fig5}(i) and \ref{fig6}(i)).

IRIS sit-and-stare observations clearly captured multiple ribbons, providing crucial diagnostics of magnetic reconnection signatures in the lower atmosphere. At about 21:13 UT, the hook structure remained positioned to the right of the slit (see Figure \ref{fig6}(a)), confirming the MFR's north footpoint did not traverse the slit during the quasi-static phase. Meanwhile, a distinct flare ribbon R3 (also visible in Figure \ref{fig3}(d3)) appeared north of the hook, exhibiting a strong enhancement along the slit at \siiv 1402.77 {\AA} line, as shown in Figure \ref{fig6}(d). We further selected three \siiv profiles before and after the appearance of R3 to present the spectral characteristics, as shown in Figure \ref{fig6}(g). It was found that the centroids of \siiv lines show a obvious redshift with the velocity at 16$\sim$18 km s$^{-1}$ at the ribbon R3, and the peak intensity enhancements up to more than 10-fold. Notably, the peak intensity of \siiv profiles at about 21:13 UT was obviously larger than that during the impulsive phase, indicating the ongoing magnetic reconnection which occured in the background field was mainly taking place in the quasi-static phase.

During the impulsive phase, the hook structure swept across the slit (see Figure \ref{fig6}(b)), with \siiv profiles showing three distinct enhancements in red-wing (between green and red dashed lines in Figure \ref{fig6}(e)), corresponding to brightenings at ribbon R3 and the outer/inner edge of the hook, respectively. The redshift velocity at the inner edge (red point in Figure \ref{fig6}(e)) was estimated to be 10 km s$^{-1}$ at about 21:20 UT, which increased to be 23 km s$^{-1}$ at about 21:22 UT, as shown in Figure \ref{fig6}(h). And the peak intensity at the inner edge of the hook significantly exceeded that of ribbon R3, indicating more violent magnetic reconnection taking place between the MFR and the nearby sheared arcades. Additionally, an evident blueshift with a velocity at 36$\sim$85 km s$^{-1}$ was detected below the erupted MFR ([X, Y] $\approx$ [576$^{\prime\prime}$, -155$^{\prime\prime}$]), where a H$\alpha$ filament (see Figure \ref{fig5}(h)) remained post-eruption, as shown in Figures \ref{fig6}(e) and \ref{fig6}(f). It was dudeced that the remained filament was substantially disturbed by the footpoint drifting behaviors of the erupted MFR, and the blueshifts was resulted from both the underlying filament perturbation and the upward motions of the erupted MFR.

\subsection{Photospheric magnetic motion and three-dimensional coronal dome}
\begin{figure*}[ht!]
	\centering 
	\includegraphics[width=0.87\textwidth,height=1.16\textwidth]{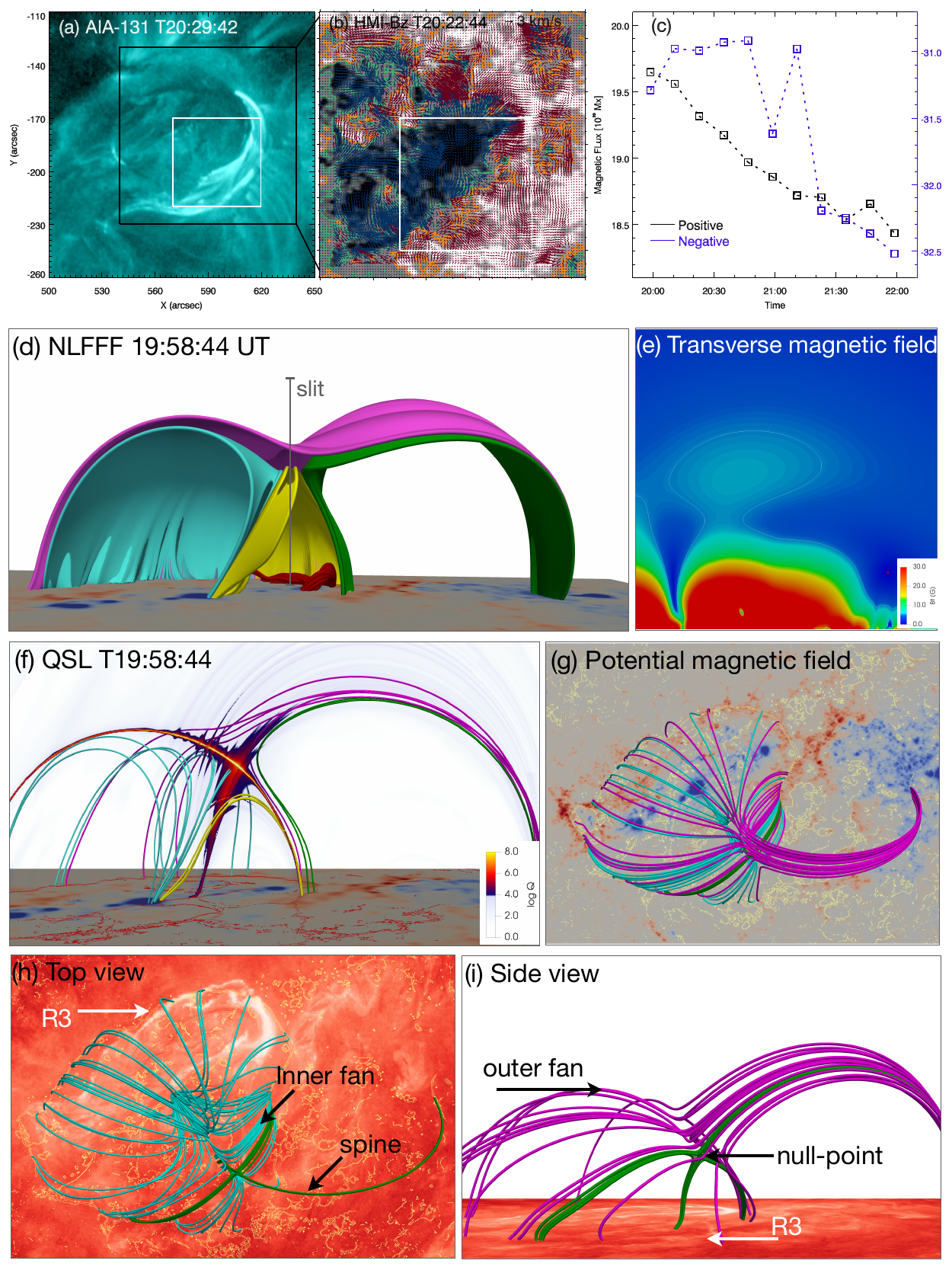}
	\caption{Magnetic features in the active region. (a) AIA 131 {\AA} image showing the filament position. (b) Bz component of the vector magnetograms superimposed over the velocity fields. The blue (red) arrows represent the velocity of the negative (positive) polarities. The green (orange) contours represent the negative (positive) magnetic polarities of $\pm$ 50 G. (c) Temporal evolution of integrated magnetic fluxes in the white box in panel (a) and (b). (d) Three-dimensional nonlinear force-free field showing a fan-spine structure. (e) The transverse magnetic field along the slit direction in panel (d). (f) Distribution of Q values in a perpendicular slice. (g) Three-dimensional potential field showing the fan-spine structure. (h) Top view of the fan-spine structure showing the inner fan and spine. The yellow contour indicates the distribution of Q values on the photosphere. The overlaid AIA 304 {\AA} image shows the circle ribbon R3 and a straight one R4. (i) Side view of the fan-spine structure showing the outer fan and null-point. \label{fig7}}
\end{figure*}

The filament eruption is deduced to be driven by the photospheric shearing and converging flows beneath the filament. Figures \ref{fig7}(b) presents the radial component (Bz) of vector magnetic fields on the pohtosphere, with the velocity field derived by Differential Affine Velocity Estimator for Vector Magnetograms method (DAVE4VM; \citealp{Schuck_2008_ApJ}). According to the direction of the blue and red arrows in the white box region, some shearing and converging motions were observed along the PILs under the filament. To confirm such an interaction, we inspected the evolution of magnetic flux within the white box region, which contained main straight part of the ribbon R1 and R2. It is found that the positive flux revealed a constant decrease during the filament eruption, with an average flux cancellation rate of $2.2 \times 10^{16}$ Mx s$^{-1}$ prior to 21:12 UT (see Figure \ref{fig7}(c)). Subsequently, the cancellation rate reduced to $8.7 \times 10^{15}$ Mx s$^{-1}$, suggesting photospheric flux cancellation as the main trigger mechanism of this filament eruption. Simultaneously, the negative flux also revealed a slight decrease during the quasi-static phase, consistent with the flux cancellation. However, the negative flux shows a significant increase after 21:12 UT, which was mainly attributable to a significant influx of negative magnetic polarities from left of the white box region.

Using the magnetofrictional technique \citep{Guoyang_2016_ApJ_2, Guoyang_2016_ApJ_1}, we reconstructed the nonlinear force-free field (NLFFF) in this AR, as shown in Figure \ref{fig7}(d). The magnetic configuration was comfired to be a fan-spine structure with the filament embedded under the dome as predicted. Moreover, taking a vertical slit perpendicular to the filament, we found that there exits an interesting region with higher transverse magnetic field above the dome as shown in Figure \ref{fig7}(e), which might be the main reason for the occurrence of the prolonged initial process. As the filament rises, the overlying magnetic field became compressed and accumulated, which in turn suppressed the further ascent of the filament. Eventually, the magnetic reconnection took place between the oppositely connected flux systems (yellow and purple lines) within the QSL, as indicated by the high squashing factor (Q) in Figure \ref{fig7}(f), reducing the constraints of background field and facilitating the subsequent eruption, analogous to the breakout magnetic reconnection model \citep{Antiochos_1999_ApJ_1}. To better present the clear fan-spine structure, We made an another potential extrapolated with a larger FOV, as shown in Figure \ref{fig7}(g)-\ref{fig7}(i). It was found that the QSL footprints along with the fan is highly accordant with the circle ribbon R3. With the MFR gradually rising up, more arched loops above the south filament would reconnect with the overlying background field in the QSL associated with the null-point, resulting in a more complete-circle ribbon R3 and a remote ribbon R4 coincide with the QSL on the lower atmosphere, as shown in Figure \ref{fig7}(h).

\section{Conclusions and discussions} \label{sec4}
In this study, we present a detailed analysis of the initial eruption process including two accelaration stages of a filament located in a complex active region, where the positive magnetic polarity was fully surrounded by negative polarities. The first acceleration stage of the eruption is supposed starting with the filament splitting motion, which was argued to be triggered by the magnetic reconnection within the filament. It was found that the upper branch (filament F1) was rapidly heated to approximately 8 MK forming a hot channel, and then slowly rised at an average speed of 22 km s$^{-1}$ under the continued magnetic reconnection resulting in the first M1.0 class solar flare. Simultaneously, the filament northern footpoint (FP2) evolved from bright blocks into a semi-enclosed hook structure, indicating the growth of pre-eruptive MFR with an estimated axial magnetic flux of $0.6 \times 10^{20}$ Mx. The second acceleration stage is associated with the rapid rise of the hot channel at about 30 minutes later, during which the axial flux of the MFR increased rapidly to over $1.1 \times 10^{21}$ Mx with an average rate of $5.7 \times 10^{17}$ Mx s$^{-1}$. As a distinct annular flare ribbon appearing outside the hook structure in IRIS/SJI 1400 images, the hot channel was rapidly accelerated to a speed exceeding 50 km s$^{-1}$ within several minutes, and went through an impulsive eruption in the subsequent ten minutes leading to an intense M4.6 class flare and a halo CME event. During this process, the 171 {\AA} images reveal that the coronal loops above the south hot channel were successively opened and underwent significant changes in their orientation, indicating the occurrence of magnetic reconnection in the background field. The annular ribbon evolved into a large circle ribbon, and the MFR$'$ footpoints (FP2) indicated by the hook structure rapidly drifted along the inner side of the circle ribbon until the obvious dimming appeared within the hook. Furthermore, the NLFFF and potential extrapolation based on the photospheric vector magnetogram confirms that the active region possesses a large scale fan-spine magnetic structure, with the filament situated beneath the fan dome. These observations declare that the filament eruption in a fan-spine structure tends to possess a prolonged initial process as a result of the constraining effect of the overlying magnetic dome. 

Distinct from the filament initial eruption in a magnetic bipole which usually includes a single coherent slow-rise phase with several to tens minutes \citep{Chengxin_2023_ApJL, Xingchen_2024_ApJ}, we identified two sequential rising phases in this event that driven by different magnetic reconnection taking place within and above the filament. Between these two rising stages, the hot channel came to a quasi-static plateau lasting approximately half an hour. The pre-erutive MFR core gradually grew up during the first rising stage, with its axial flux increased from $0.6 \times 10^{20}$ Mx to $3.7 \times 10^{20}$ Mx at an average rate of $1.4 \times 10^{17}$ Mx s$^{-1}$. Therefore, we suggest that the quasi-static plateau is a critical process for the successful filament eruption beneath a magnetic dome, as it allows continuous growth of the flux rope until it can overcome the confinement imposed by the dome. Moreover, regarding the evolution from filament split to impulsive eruption as a prolonged slow rise phase that lasted more than one hour, the separation of two flare ribbons shows a strong correlation with the kinematic evolution of the flux rope, with the ribbons separated at a velocity of 1.2 km s$^{-1}$ and 4.8 km s$^{-1}$ during the slow rise and second rising phase, respectively.

By comparing the photospheric magnetic field distribution beneath the dome, we find that the positive polarity corresponding to the outer fan is comparable in magnetic strength to the negative polarity field corresponding to the inner fan in this event (see Figure \ref{fig1}(d)). This configuration differs significantly from the magnetic distribution beneath the filament reported by \cite{Joshi_2017_ApJ} (see their Figure 1(b), where the positive field is noticeably weaker than the surrounded negative field). Hence, a more robust dome structure overlies the filament in this case. Whenever the filament begins to rise, whether due to magnetic reconnection or an ideal instability, the MFR will be obstructed owing to the flux pile-up in its front, which is likely the primary mechanism leading to the observed two-stage eruption \citep{Byrne_2014_SoPh, Xuzhe_2024_MNRAS}. Moreover, when a strong transverse field is present above the dome, as shown in Figure \ref{fig7}(e), it would further restrain the MFR’s ascent, resulting in an extended quasi-static phase. Additionally, the magnetic reconnection occurring in the QSL at the dome may be a crucial point for the eruption: when the reconnection is dominated by the opposite background fields on either side of the QSL, it tends to reduce the overlying confinement, allowing a successful eruption \citep{Byrne_2014_SoPh, Reeves_2015_ApJ}; whereas if the reconnection occurs between the background field and the MFR, it may potentially erode the main body of the MFR and lead to a failed eruption \citep{Chenjun_2023_ApJL, Xuzhe_2024_MNRAS}.

On the other hand, the southern leg of the hot channel expanded rapidly outward while the northern leg remained steady after the first accelaration stage. We speculate that this asymmetry motion is related to stronger coronal magnetic fields above the filament northern foot, as the sunspot group was predominantly located north of the filament. Additionally, the birghtening points enveloping the MFR southern footpoint that anchored within the sunspot umbra largely retains its original shape during the slow rise phase, whereas the northern footpoint, rooted in a weak-field region, expands quickly and forms a larger hook structure. It suggests that the footpoints embedded in weak magnetic fields behave as relatively free ends, thereby promoting footpoint drifts and facilitating the release of free energy stored in the flux rope. In contrast, the fixed footpoints in strong-field regions restrict energy release, leading to deformation in the leg above such anchored footpoints.

\begin{acknowledgments}
Thanks the editors and reviewers$'$ comments on our manuscript, which are highly valuable and helpful for revising and improving this paper, as well as the important guiding significance to our researches. We appreciate the SDO/AIA, SDO/HMI, IRIS and CHASE teams for providing the high-quality observational data. CHASE mission is supported by China National Space Administration. And we thanks Dr. J. Chen for his assistance on analysing the coronal magentic structure. This project is supported simultaneously by the Fundamental Research Funds for the Central Universities, Sichuan Science and Technology Program (2025ZNSFSC0877), and the National Natural Science Foundations of China (NSFC 12373063).
\end{acknowledgments}

\appendix
Figure \ref{afig1} shows the filament splitting process prior to the slow rise phase observed in AIA 171 {\AA} and 304 {\AA}. At about 19:13 UT, the filament remained stable with an overall inverse C-shaped. Subsequently, at about 19:43 UT, a brightening along the filament threads was observed within the filament, as indicated by the blue arrows in Figure \ref{afig1}(b) and \ref{afig1}(h), suggesting the possible occurrence of magnetic reconnection inside the filament. With more brightenings appeared, the south filament was gradually split and separated into two distinct sections at around 20:20 UT: the upper section F1 and the lower section F2, as indicated by the black and white arrows. Meanwhile, as the internal brightening developed, an inverse C-shaped bright loop intermittently appeared in 94 {\AA} marked by the yellow dashed line, as shown in Figure \ref{afig2}, suggesting the formation and growth of the pre-eruptive hot channel through the potential tether-cutting reconnection. By approximately 20:45 UT, a clearly coherent hot channel was formed, consistent with the structure in AIA 131 {\AA} shown in Figure \ref{fig2}(a2).

\renewcommand{\thefigure}{A\arabic{figure}}
\setcounter{figure}{0}
\begin{figure*}[ht!]
	\centering 
	\includegraphics[width=0.9\textwidth,height=0.69\textwidth]{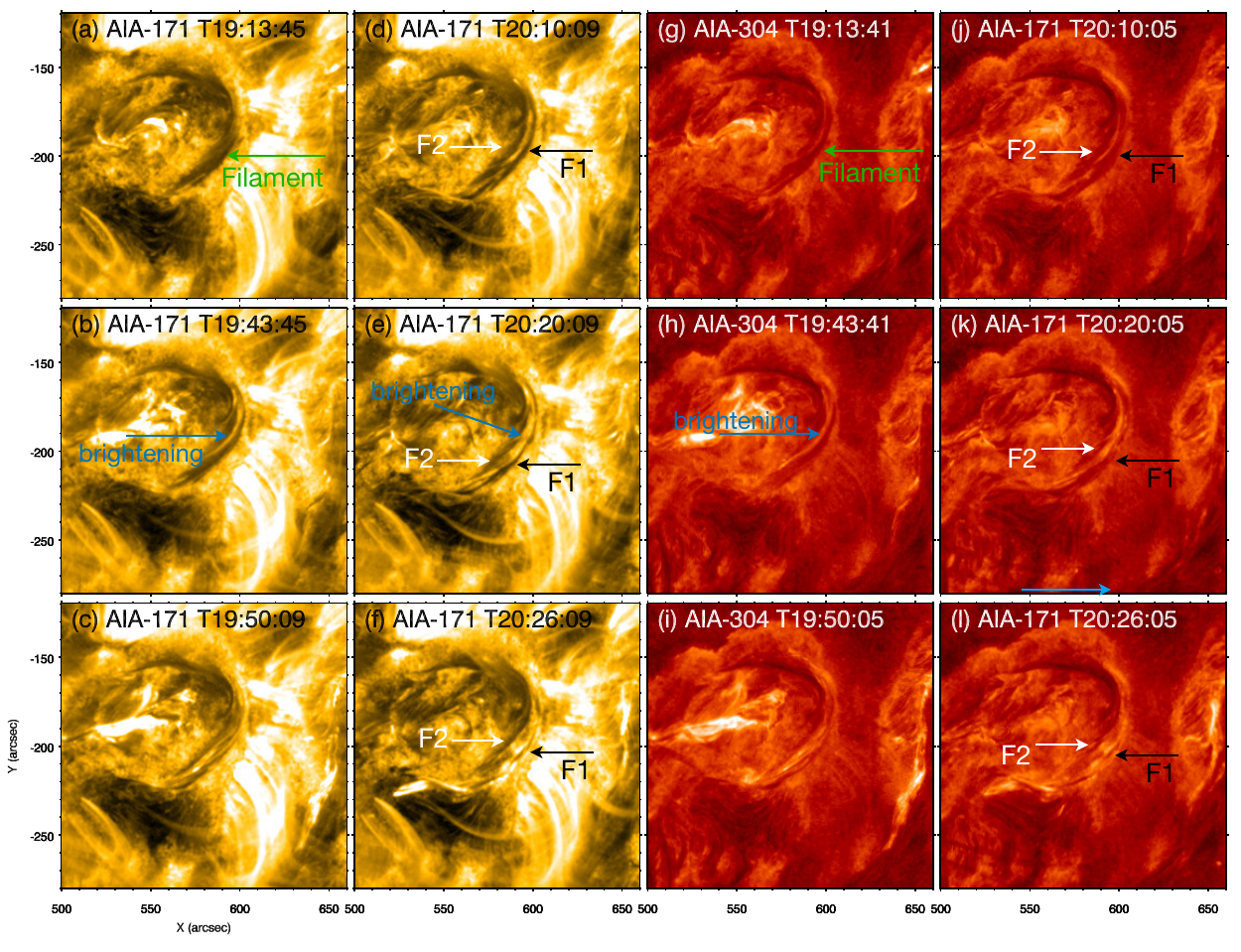}
	\caption{Filament splitting process prior to the eruption in AIA 171 {\AA} ((a)-(f)) and 304 {\AA} ((g)-(l)). The green arrows indicate the south filament before splitting. The blue arrows indicate the brightening within the filament. The black and white arrows indicate the upper (F1) and lower (F2) section. A 17 s animation including AIA 94 {\AA} and 171 {\AA} is available online to show the hot channel formation and eruption from 19:00 UT to 22:30 UT. And a color inversion in AIA 171 {\AA} was applied to better highlight the evolution of the overlying coronal loops. \label{afig1}}
\end{figure*} 

\begin{figure*}[ht!]
	\centering 
	\includegraphics[width=0.9\textwidth,height=0.48\textwidth]{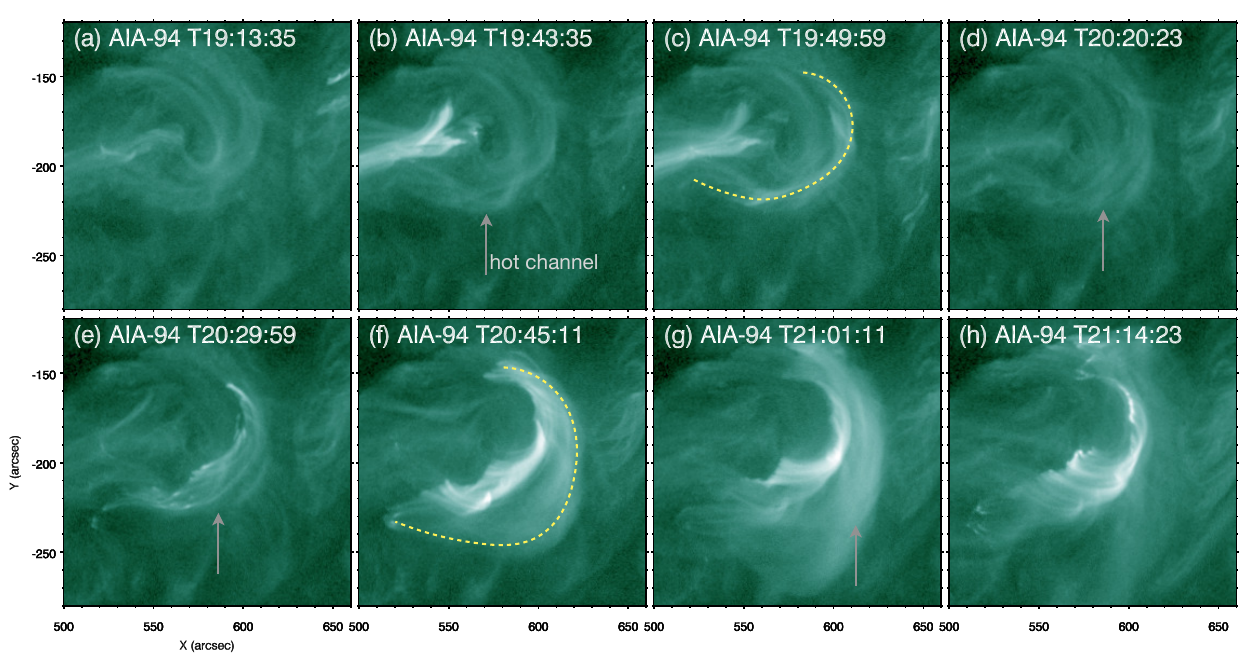}
	\caption{Formation process of the hot channel in 94 {\AA}. The yellow dashed lines indicate the inverse C-shaped bright loop, representing the observational hot channel. \label{afig2}}
\end{figure*}

\bibliography{refer}{}
\bibliographystyle{aasjournalv7}

\end{document}